\documentclass[jap,twocolumn]{revtex4}
\usepackage[dvips]{graphicx}
\usepackage{amsmath}
\usepackage{bm}
\usepackage{color}

\begin{document}
\title{Simulation of current-induced microwave oscillation \\in geometrically confined domain wall.}

\author{Katsuyoshi Matsushita}
\author{Jun Sato}
\author{Hiroshi Imamura}

\affiliation{Nanotechnology Research Institute (NRI), Advanced
Industrial Science and Technology (AIST), AIST Tsukuba Central 2,
Tsukuba, Ibaraki 305-8568, Japan.}

  \begin{abstract}
   We studied magnetization dynamics of a geometrically confined
   domain wall under dc current by solving simultaneously  the
   Landau-Lifshitz-Gilbert equation and diffusion equation for spin
   accumulation. 
   We showed that the oscillation motion of the domain wall is driven 
   by the spin-transfer torque and the dc current is converted to the ac
   voltage signal. 
   The results means that the geometrically
   confined domain wall is applicable as a source of microwave oscillator.
  \end{abstract}
  \maketitle


  Achievement of a nano spin-transfer oscillator is one of important
  issues in microwave technology.  Many studies have succeeded in
  development of microwave oscillation in nanopillers and point
  contacts\cite{Katine:2000,Tsoi:2000,Kiselev:2003,Rippard:2004,Covington:2004,Krivorotov:2004,Kasa:2005,Mancoff:2005}.
  Recently possibilities of another kind of nano spin-transfer
  oscillators using a domain wall were proposed\cite{He:2007,Ono:2008}.
  In those works, the oscillation generating microwave in GMR or TMR
  devices is not an uniform oscillation of the free layer magnetization
  but an oscillation\cite{He:2007} or magnetization rotational motion\cite{Ono:2008} of
  the domain wall.
  
  On the other hand, it is known that the resistance due to the domain
  wall depends on its magnetic structure\cite{Levy:1997,Brataas:1999,Simanek:2001}.
  If the oscillation motion of the magnetic structure of the domain wall
  is induced by spin transfer torque under dc current, 
  we can obtain the ac voltage signal and can use the domain wall as a
  microwave source.  
  
  
  In this paper, we numerically calculate the dynamics of the
  geometrically confined domain wall \cite{Bruno:1999} under the dc
  current by solving simultaneously the Landau-Lifshitz-Gilbert equation
  and diffusion equation for spin accumulation. 
  We show that the oscillation of the magnetization vector of the domain
  wall is induced by the spin-transfer torque and we can obtain the ac
  voltage signal due to the domain wall oscillation.
  We also show that the power spectrum of the MR ratio has a sharp peak
  corresponding to the oscillation of the domain wall.
  The frequency of the peak is about 1 hundred GHz for the current density
  of about 0.01 mA/nm$^2$ which is reachable value in experiments with avoiding breakdown by Joule heating.

\begin{figure}[tb]
    \begin{center}
     \includegraphics[scale=0.6]{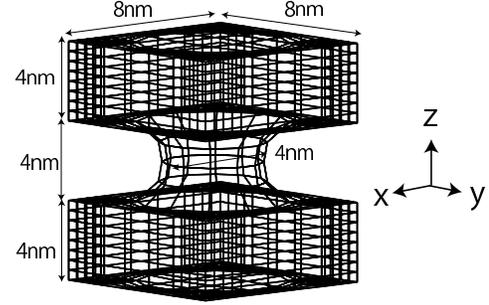}
    \end{center}
    \caption{The model of a geometrically confined domain wall for a finite element method.}
    \label{fig:model}
   \end{figure}

  The system, we consider, is shown in Fig.~\ref{fig:model}.
   The system consists of two magnetic electrode layers and a
   non-magnetic insulator layer sandwiched by the electrodes.  The
   insulator layer contains magnetic bridge which geometrically confines
   a domain wall.  The system is divided into about 1500 hexahedral
   finite elements, where spin accumulation and demagnetization field
   are evaluated.  The size of magnetic metal electrode at the bottom
   and top layers is 8 $\times$ 8 $\times$ 4 nm.  The shape of the
   bridge between two electrodes is a rotated elliptic arch around the
   center axis perpendicular to the layers, the diameters of the bottom
   and center layers of which are set at 6 and 4 nm, respectively.  
   The size of the system is larger than the diffusion length of 2nm of CoFe\cite{Moyerman:2005} at room temperatures.
   Thus we can deal with the electron system in diffusive limit, unlike a previous ballistic treatment\cite{ohe:2006}.
   Then as previously predicted by $\breve{\rm S}$im\'{a}nek\cite{Simanek:2001},
   the spin accumulation mainly induces domain wall resistance.
  
   The magnetization vectors are expressed by the classical spins on
   the simple cubic lattice with the lattice constant of $a= 0.4$nm.  
   The Hamiltonian ${\cal H}$ is given by
   \begin{eqnarray}
    &&{\cal H} = -J_{\rm dd} \sum_{\left<i,j\right>} \vec S_{i} \cdot \vec
    S_{j} + J_{\rm sd} \sum_i \vec S_{i} \cdot \delta \vec m_i 
    \nonumber \\ 
    \mbox{}&&\hspace{0em}
    + \frac{K_d}{4\pi} \sum_{i} \vec S_{i} \cdot \int d\vec r 
    \left\{ \frac{\hat 1}{|\vec r_{i}|^3}-3\frac{\vec r_{i} \otimes \vec r_{i}}{|\vec r_{i}|^5} \right\} \cdot \vec S(\vec r),
    \label{Hamiltonian}
   \end{eqnarray}
   where $\vec{r}_{i}$ represents the relative coordinate of the $i$-th site from the position $\vec r$, $\vec
   S_{i}$, the classical Heisenberg spin with absolute value of unity and,
   $\delta \vec m_{i}$, local spin accumulation
   density at the $i$-th site.  
   Such the semi-classical treatment is justified for the system with ferromagnetic order
   because the magnetization is larger than its quantum fluctuation.
   The first
   and second terms in the right hand side of Eq.~\eqref{Hamiltonian}
   express the exchange interactions between local magnetizations at
   nearest neighbor sites, and between local magnetization and spin
   accumulation at each site, respectively.  The exchange coupling
   constant between nearest local magnetization denoted by $J_{\rm dd}$
   is related to the exchange stiffness constant in continuous limit,
   $A$, and a lattice constant, $a$, with $J_{\rm dd} = 2aA$.  We set
   $J_{\rm dd}$ at 0.04 eV which is on the order of the transition
   temperature of CoFe, $T_c \sim 1.2 \times 10^3 K$.   The exchange
   coupling constant $J_{\rm sd}$ is set at 0.1 eV in accordance with
   Ref.\cite{Zhang:2002}.  
   The third term denotes the dipole-dipole interaction energy.  
   In order to deal with the dipole-dipole interaction in a complex
   system shape, we adopt a finite element - boundary element (FEM-BEM)
   hybrid method~\cite{Fredkin:1990} on the spin field on the continuum space,
   $\vec S(\vec r)$, which is defined for each position, $\vec r$, by the
   spin field extrapolated from the lattice sites to the position.  The
   exchange length for the dipole-dipole interaction is $l_{\rm ex}
   =\sqrt{J_{\rm dd}/2K_d} \sim3$ nm, which is comparable to the length of
   the considered confining region of 4nm. Thus
   the thickness of the domain wall depends on both the size of the
   confining region and the exchange length.
   
   In the adiabatic approximation, the local spin accumulation density,
   $\delta \vec m_{i}$, is determined by solving the 
   following differential equations\cite{Zhang:2002},
   \begin{align}
    &\frac{\partial}{\partial t} \delta \vec m(\vec r) =
    \nabla \left\{ \beta \vec S(\vec r) \vec j_{\rm e}(\vec r)  
    +\hat A(\vec S(\vec r)) \delta \vec m(\vec r)  \right\} 
    \label{eq:ZLF1}\nonumber\\
    &\ \hspace{1.3cm}
    + \frac{J_{\rm sd}}{\hbar}\delta \vec m(\vec r)
    \times \vec S(\vec r) + \frac{\delta \vec m(\vec r)}{\tau},\\
    &\hat A(\vec S(\vec r)) = 2D_0\left[\hat 1 - \beta^2\vec S(\vec r)
    \otimes \vec S(\vec r)\nabla\right], 
    \label{eq:ZLF2} 
   \end{align}
   where $\vec j_e$, $C_0$, $D_0$, $\beta$ and $\tau$ denote electronic current, 
   conductivity, diffusion constant, polarization of resistivity and 
   relaxation time due to a spin-orbit interaction respectively. 
   Equations~\eqref{eq:ZLF1}-\eqref{eq:ZLF2} are solved numerically
   with combining continuous equation for electronic current, 
   $\nabla \vec j_e = 0 \label{eq:continous}$. 
   $C_0$ and $\beta$ are taken to be those for conventional ferromagnets1
   as $C_0 =$70$\Omega$nm and $\beta=0.65$, respectively. $D_0$ and
   $\tau$ are obtained, respectively, by the Einstein relation, $C_0 =
   2e^2N_{\rm F}D_0$ and $\lambda = \sqrt{2\tau D_0(1-\beta^2)}$ for given diffusion
   length $\lambda$, density of states at the Fermi level, $N_{\rm F}$
   and electron charge $e$. Here we employ $\lambda$=12nm and $N_{\rm
   F}=$7.5nm$^{-3}$eV$^{-1}$.
   For the boundary condition of the spin accumulation, we artificially adopt
   $\delta \vec m = \vec 0$ at the top and bottom layers,
   ignoring any parasitic resistance.
   On the other boundaries, the natural boundary condition is employed.
   In the case,
   the simulated spin accumulation distribution is nonuniform and concentrate 
   on the contact region.
   
   Substituting the solution of Eqs.~\eqref{eq:ZLF1} and \eqref{eq:ZLF2}
   into Eq.~\eqref{Hamiltonian}, we evaluate the Hamiltonian of
   Eq.~\eqref{Hamiltonian}.  The effective magnetic field for
   $\vec{S}_{i}$ is given by $\partial {\cal H}/\partial \vec S_i$ and 
   the dynamics of $\vec{S}_{i}$ is determined
   by the following Landau-Lifshitz-Gilbert equation,
   \begin{equation}
    \frac{d}{dt} \vec S_i 
     = \frac{\gamma}{1+{\alpha}^2} \vec S_i \times
     \left(  \frac{\partial {\cal H}}{\partial \vec S_i} + \alpha \vec
      S_i \times \frac{\partial {\cal H}}{\partial \vec S_i}\right),
   \end{equation}
   were $\gamma$ and $\alpha$ denote the gyromagnetic ratio and the
   Gilbert damping constant.  The equation is numerically solved by a
   quaternion method\cite{Visscher:2002}.  The time step $\Delta t$ and
   $\alpha$ are set at 3.4 $\times$ 10$^{-2}$ fs and 0.02,
   respectively. The number of sites, $N_{\rm s}$ is about $10^4$.  
   For simplicity, we use antiparallel boundary condition of the
   spins in order to simulate the experimental situation with a
   180$^\circ$ domain wall.  On the boundary, $\vec S$'s are fixed at
   (-1,0,0) on the top electrode and (+1,0,0) on the bottom electrode
   except for the boundary surface between the magnetic and insulator
   layers. The directions of spins, $x$, $y$ and $z$ are defined as shown
   in Fig.~\ref{fig:model}.

   \begin{figure}[t]
    \begin{center}
     \includegraphics[scale=0.7]{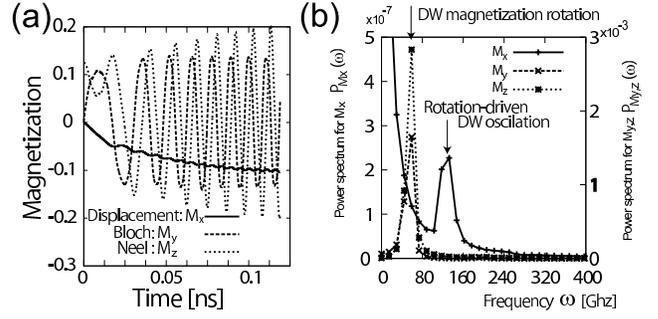}
    \end{center}
    \caption{(a) The $x$, $y$, and $z$-components of the mean
    magnetization vector, $\vec M$,  is plotted as a function of time
    $t$ for $j=0.012$mA/nm$^2$(b) The power spectra of the magnetization, $P_{x,y,z}$, are plotted as a
    function of frequency $\omega$.  }
    \label{fig:Magnetization}
   \end{figure}
   
   In the present work, we employ current of 0.012 mA/nm$^2$. The current is applied in
   the perpendicular direction to the layers.  The simulation starts
   from stable N\'{e}el wall state obtained by a simulation without dc
   current density.
   
   In order to understand dynamics of the local magnetization distribution of the domain
   wall, we calculate the magnetization $\vec M
   = \left(\sum_i \vec S_i\right)/N_{\rm s}$ for the whole of the system.
   The typical result in time dependence of the magnetization
   is shown in Fig.~\ref{fig:Magnetization}(a).  We observe
   steady regular rotation around the $x$-axis. 
   Such rotation is known for the moving domain wall in
   wire systems above Walker's threshold current\cite{Tatara:2004}.  The
   magnetic structures with $M_y$=0 and $M_z$=0 correspond to  the
   N\'{e}el wall and the Bloch wall, respectively.  Thus the rotation
   corresponds to an oscillation between the N\'{e}el and Bloch walls.
   On the other hand, the displacement of the domain wall center is
   characterized by the value of $M_x$ because the magnetization vectors
   at the boundary surface are aligned to be parallel to the $x$-axis.
   As shown in Fig.\ref{fig:Magnetization}(a)  that the domain
   wall displacement oscillates as a function of time.
   
   Figure~\ref{fig:Magnetization}(b) shows the power spectra of the
   magnetization motions, which are defined by
   \begin{equation}
    P_{\rm x,y,z}(\omega) 
    = \frac{1}{T}\int\hspace{-0.75em} \int_{T_0}^{T_0+T} \hspace{-1em} e^{-i\omega
     t}
     M_{\rm x,y,z}(\tau)M_{\rm
     x,y,z}(t-\tau) d\tau dt, 
   \end{equation}
   where $T_0$ and $T$ are much longer than the periods of the
   oscillations. Each spectrum has a single characteristic
   frequency. The frequency of the displacement oscillation is twice of that of the rotational motion.
   This is because the displacement oscillation originates from coupling with the rotational motion driven by the spin transfer torque. For the N\'{e}el and Bloch walls, respectively, the displacement takes minimum and maximum.

\begin{figure}[t]
    \begin{center}
     \includegraphics[width=0.8\columnwidth]{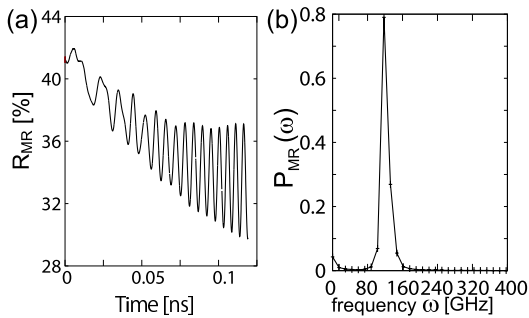}
    \end{center}
    \caption{(a) The magnetoresistance ratio $R_{\rm MR}$ is plotted as
    a function of time $t$ for 0.01mA/nm$^2$. (b) 
    The power spectrum of the magnetoresistance ratio $P_{\rm MR}$ is
    plotted as a function of frequency $\omega$. }
    \label{fig:MR}
   \end{figure}

   The oscillation of the magnetization vectors induces the oscillation
   of the spin accumulation through Eq.~\eqref{eq:ZLF1}.
   The electric field at position $\vec{r}$ is given by
   \begin{equation}
    \vec E(\vec r) = \frac{1}{2C_0}
     \left(\vec j_{\rm e}(\vec r)+
      2D_0\left[\hat 1 + \beta \vec \sigma \cdot \vec S(\vec r)\right]
      \delta \vec m(\vec r)\right).
     \label{eq:voltage}
   \end{equation}
   The voltage drop of the system is obtained by integrating the
   electric field $\vec{E}(\vec{r})$ along $z$-axis.
   From Eq.~\eqref{eq:voltage}, the oscillation in the spin accumulation
   then yields the oscillation in magnetoresistance ratio, $R_{\rm
   MR}(t)=R(t)/R_0-1$, where $R_0$ and $R(t)$ denote resistance with and
   without a domain wall, respectively. $R_{\rm MR}$ oscillates as a
   function of time $t$ as shown in Fig.~\ref{fig:MR}(a).  The power
   spectrum of $R_{\rm MR}$ defined by
   \begin{equation}
    P_{\rm MR}(\omega)
     = \frac{1}{T}\int\hspace{-0.75em}
     \int_{T_0}^{T_0+T} \hspace{-1em} 
     e^{-i\omega t}
     R_{\rm MR}(\tau)R_{\rm MR}(t-\tau) d\tau dt
   \end{equation}
   is shown in Fig.~\ref{fig:MR}(b). In the accuracy of our calculation,
   $P_{\rm MR}$ has a characteristic isolated peak.  Thus the dc current
   is converted to the ac voltage signal with a characteristic frequency.
   The frequency of the peak agrees with that of the $P_{x}$
   denoted by  the arrow in Fig~\ref{fig:Magnetization}(b).
   Thus the MR oscillation is due to periodically change of the spin accumulation induced by the displacement oscillation. 
    
   We note that only the resistance due to the spin accumulation is
   considered.  However in the real system the resistance originates
   from not only the spin accumulation but also scattering of electrons
   due to the domain wall\cite{Levy:1997,Brataas:1999}.  The correction
   will enhance the magnetoresistance ratio and therefore the power of
   the microwave oscillation as compared with the results of our
   simulation.
   

   In conclusion, we studied the magnetization dynamics of the 
   geometrically confined domain wall under dc current by solving the
   Landau-Lifshitz-Gilbert equation and diffusion equation for the
   spin accumulation.  
   From calculation result, we propose a scenario of the microwave generation
   as follows: current induced spin-transfer torque drives magnetization rotation in the domain wall.
   Then the displacement oscillation of domain wall is induced by the coupling with the rotation and drives spin accumulation oscillation. As a result the microwave oscillation in voltage signal appears.
   We conclude that the geometrically confined domain wall is
   applicable as a source of microwave generator.

   The authors would like to thank M.~Doi, H.~Iwasaki, M.~Ichimura, M.~Takagishi,
   M.~Sahashi, M.~Sasaki, T.~Taniguchi, N.~Yokoshi and K.~Seki for useful discussions.  The work was supported by The New Energy and Industrial Technology Development
   Organization (NEDO).  K.~M. was supported by a Grant-in-Aid for Young
   Scientists from the Ministry of Education, Science, Sports and Culture
   of Japan.
   

\end{document}